\documentclass[useAMS,usenatbib]{mn2e}

\usepackage[latin1]{inputenc}
\usepackage{graphicx}
\usepackage{epstopdf}

\usepackage{rotate}
\usepackage{amsmath, float}
\usepackage{amssymb}
\usepackage{soul}
\usepackage{txfonts}
\usepackage{fleqn}
\usepackage{color}
\usepackage{comment}

\usepackage{mathrsfs}  
\usepackage{booktabs}
\usepackage{colortbl} 

\newcolumntype{G}{>{\columncolor[gray]{0.8}}l} 

\newcommand{\be}{\begin{equation}}
\newcommand{\ee}{\end{equation}}
\newcommand{\bdm}{\begin{displaymath}}
\newcommand{\edm}{\end{displaymath}}
\newcommand{\bea}{\begin{multline}}
\newcommand{\eea}{\end{multline}}
\newcommand{\ba}{\begin{align}}
\newcommand{\ea}{\end{align}}

\newcommand{\ltsim}{\mbox{{\raisebox{-0.4ex}{$\stackrel{<}{{\scriptstyle\sim}}$}}}}
\def\simlt{\mathrel{\hbox{\rlap{\hbox{\lower4pt\hbox{$\sim$}}}\hbox{$<$}}}}
\def\simgt{\mathrel{\hbox{\rlap{\hbox{\lower4pt\hbox{$\sim$}}}\hbox{$>$}}}}

\title[Radio Circular Polarization in the Crab Nebula]
{Modeling Radio Circular Polarization in the Crab Nebula}
\author[N. Bucciantini, B. Olmi]{
N. Bucciantini$^{1,2,3}$\thanks{E-mail: niccolo@arcetri.astro.it},
 B. Olmi$^{2,1,3}$ \\
$^{1}$INAF - Osservatorio Astrofisico di Arcetri, Largo E. Fermi 5,
I-50125 Firenze, Italy\\
$^{2}$Dipartimento di Fisica e Astronomia, Universit\`a degli Studi di Firenze, Via G. Sansone 1, 
I-50019 Sesto F.~no  (Firenze), Italy\\
$^{3}$INFN - Sezione di Firenze, Via G. Sansone 1, I-50019 Sesto F.~no  (Firenze), Italy}

\begin{document}
 
\date{Accepted / Received}

\maketitle

\label{firstpage}

\begin{abstract}
In this paper we present, for the first time, simulated maps of the
circularly polarized synchrotron emission from the Crab nebula, using
multidimensional  state of the art models for the magnetic field geometry. Synchrotron emission is the signature  of non-thermal
emitting particles, typical of many high-energy astrophysical sources,
both Galactic and extra-galactic ones. Its spectral and polarization
properties allow us to infer key informations on the particles
distribution function and magnetic field geometry. In recent years our
understanding of pulsar wind nebulae has improved substantially thanks
to a combination of observations and numerical models. A robust detection or
non-detection of circular polarization will enable us to discriminate
between an electron-proton plasma and a pair plasma, clarifying once
for all  the origin of the radio emitting particles, setting
strong constraints on the pair production in pulsar
magnetosphere, and the role of turbulence in the nebula. 
Previous attempts at measuring  the circular polarization have only provided
upper limits, but the lack of accurate estimates, based on reliable
models, makes their interpretation ambiguous. We show here that those
results are above the expected values, and that current polarimetric
tecniques are not robust enough for conclusive result, suggesting that
improvements in construction and calibration of next generation radio
facilities are necessary to achieve the desired sensitivity.

\end{abstract}


\section{Introduction}

Pulsar wind nebulae (PWNe) form when the relativistic wind from a
pulsar interacts with the environment (either the ISM or the parent
supernova remnant), leading to the formation of a synchrotron emitting
bubble, that shines in a  broad range of frequencies from radio wavelengths to
$\gamma$-rays (see \citealt{Gaensler_Slane06a} for a review). Among
PWNe, the Crab nebula \citep{Hester08a} plays a special role. It is
perhaps one of the most studied object in the sky: we know its birth
date, and the spin down properties of the central pulsar
\citep{Lyne_Pritchard+93a}; we have good estimates for the mass of the progenitor
(the mass of the confining ejecta) \citep{MacAlpine_Satterfield08a}; its spectrum
as been observed from low radio frequencies \citep{Baldwin71a,Baars72a}, through mm/IR \citep{Mezger_Tuffs+86a,Bandiera_Neri+02a},
optical/UV  radiation \citep{Veron-Cetty_Woltjer93a,Hennessy_OConnell+92a}, X-rays \citep{Kuiper_Hermsen+01a} and $\gamma$-rays from 
MeV to TeV energies \citep{Aharonian_Akhperjanian+04a,Albert_Aliu+08a,Abdo_Ackermann+10a}. This system has been modeled in a variety of different ways: one-zone models \citep[e.g.][]{Pacini_Salvati73a,Gelfand_Slane+09a,Bucciantini_Arons+11a,Martin_Torres+12a}, 1D
\citep[e.g.][]{Kennel_Coroniti84a,Kennel_Coroniti84b,Blondin_Chevalier+01a,van-der-Swaluw_Achterberg+01a,Bucciantini_Blondin+03a}, 2D
axisymmetric  \citep[e.g.][]{Begelman_Li92a,Komissarov_Lyubarsky04a,Del-Zanna_Amato+04a,Bogovalov_Chechetkin+05a,Volpi_Del-Zanna+08a}, and more recently 3D \citep{Porth_Komissarov+14a,Olmi_Del-Zanna+16a}. Today there is a general
consensus on the key properties of this PWN: the strength and
geometry of the magnetic field, the particle content, the dynamics.

Despite all of this,  we still do not know the origin of the
emitting particles, especially the radio emitting ones. This is mainly 
because we do not know if the nebula is filled by an electron-positron
plasma, which will point to an origin of radio emitting particles from the pulsar \citep{Arons12a}, or simply an
electron-proton plasma, that will suggest emitting particles come
from the environment \citep{Lyutikov03a,Komissarov13a} or the
supernova \citep{Atoyan99a}. Much of the problem stems from the fact that
theoretical models of pair production in pulsar magnetospheres \citep{Hibschman_Arons01a,Takata_Wang+10a,Timokhin_Arons13a,Takata_Ng+16a}
under-predict by orders of magnitude the pair multiplicity required by
spectral modeling \citep{Bucciantini_Arons+11a}, and this is a common problem even for other
PWNe. Moreover radio emission is compatible with a uniform particle
distribution \citep{Olmi_Del-Zanna+13a}, leaving no hint on the possible injection site, as
opposite to X-ray emitting particles, that are concentrated in the central
region. The radio spectrum is too flat \citep{Bietenholz_Kassim+97a} for standard diffusive shock
acceleration \citep{Gallant_Hoshino+92a,Achterberg_Gallant+01a,Sironi_Spitkovsky11a}, which suggests a different acceleration mechanism,
possibly unrelated to the pulsar wind \citep{Zrake16a,Tanaka_Asano17a,Zhdankin_Werner+17a}.

Knowing if the relativistic plasma contains or not a large fraction
of positrons (i.e. if it is an electron-positron or electron-proton
plasma), is important for our understanding of the pulsar central
engine, and its magnetospheric activity, but it also has important
consequences for the study of relativistic engines in general
[problems of composition of relativistic outflows are present also for
extragalactic radio jets \citep{Ghisellini_Celotti+92a,Reynolds_Fabian+96a,Hirotani_Iguchi+00a,Kino_Kawakatu+12a} and GRB jets \citep{MacFadyen_Woosley99a,Popham_Woosley+99a,McKinney05a,Metzger_Giannios+11a}], for the
possible origin of the so called positron excess in cosmic rays
\citep{Adriani_Barbarino+09a,Hooper_Blasi+09a,Blasi_Amato11a,Adriani_Bazilevskaya+13a,Aguilar_Alberti+13a},
and for alternative models to dark matter annihilation in the Galactic
center \citep{Wang_Pun+06a}.

The presence of positrons could be verified looking at internal dispersive
effects in plasmas (which are strongly dependent on the mass ratio of
positive versus negative charges), but these are usually weak
compared to those due to the propagation in the ISM. On the other
hand, as it was suggested by \citet{Wilson_Weiler97a},  residual circular polarization
(CP) from
synchrotron emission could be used to constrain the presence of
positrons. Synchrotron emission by relativistic particles of Lorentz
factor $\gamma$ spiraling in
a magnetic field, inclined at an angle $\theta$ with respect to the
line of sight, is known to be elliptically polarized \citep{Legg_Westfold68a}, but
the ratio of Stokes parameters $V/I\simeq \cot{\theta}/\gamma$ is
usually so small that it is common to take it as purely linearly
polarized. A few attempts were done in the past to measure the level
of CP from the Crab nebula: \citet{Wright_Forster80a} found un upper
limit $V/I < 0.06$ at 23 GHz,
\citet{Wilson_Weiler97a} found un upper limit $V/I< 3\times 10^{-4}$
of at 610 MHz, \citet{Wiesemeyer_Thum+11a} found un upper limit $V/I < 0.002$
at 89.2 GHz. Among all of these measures, the one setting the stronger
constraint is the one by \citet{Wilson_Weiler97a}, given that the
ratio $V/I$ is smaller at higher
frequencies. Unfortunately, the interpretation of these results are
all based on such a gross model for the structure of the magnetic field
in the nebula (assumed to be uniform in strength and direction), that
the conclusions that are drawn are unreliable. For example in the
original work by \citet{Wilson_Weiler97a} it was concluded that the
measured limit on $V/I $
already indicated the presence of positrons, while in the more recent
paper by \citet{Linden15a}  it was shown that, accounting somewhat arbitrarily for
depolarization, such measure was almost an order of magnitude above the
expected threshold. These differences are due to the fact that the
geometry of the magnetic field plays an important role in determining
the amount of CP, especially for integrated measures. Even simple
recipes \citep{Burch79a,Linden15a} to tie the level of linear depolarization (which is easily
measured in the Crab nebula) to the possible
fluctuations of the magnetic field can be shown to fail if extended
trivially to CP.  It can be shown
for example that an axisymmetric field (with a uniform particle
distribution) leads to  complete vanishing CP, even if the level of
linear polarization is still a sizable fraction of the theoretical
maximum. These problems were already recognized by \citet{Linden15a}, where a
somewhat arbitrary parametrization of the depolarization effect was
introduced. 

Today, our numerical techniques enable us to compute realistic 3D models
for the structure and geometry of the magnetic field in PWNe
\citep{Porth_Komissarov+14a,Olmi_Del-Zanna+16a}, and the related
emission maps, including polarimetry. These offer for the first time the opportunity to evaluate properly
the CP emission expected from the Crab nebula if the emitting plasma
is formed by electrons-protons, and to set meaningful thresholds for
existing and future observations. Here we present for the first time a
computation of the synchrotron  CP expected from the Crab nebula,
based on the more recent 3D simulations of this object,  showing
the level of the $V/I$ ratio and its dependence  on the angular
resolution. These results are then compared with existing estimates and
observations, and discussed in the light of present day polarization
performance of low frequency radio telescopes.

This paper is organized as follows: in Sect.~\ref{sec:model} we
briefly illustrate our model and the way CP maps are computed; in
Sect.~\ref{sec:results} we present our results, and discuss them; in Sect.~\ref{sec:conclusion} we
present our conclusion with
respect of current radio facilities.

\section{The Model}
\label{sec:model}

The geometrical structure of the magnetic field is obtained from a 3D
numerical simulation in relativistic MHD, of the interaction of the
pulsar wind with the parent supernova remnant, calibrated to the case
of the Crab
nebula [\citet{Olmi_Del-Zanna+14a,Olmi_Del-Zanna+15a,Olmi_Del-Zanna+16a} to which the reader is referred for a detailed
 description of the model and input parameters]. The wind has a latitude dependent energy flux
$\sim \sin^2{\theta}$ (where $\theta$ is the colatitude with respect
to the spin axis of the pulsar), with an equator to pole anisotropy of
$\sim 10$ \citep{Spitkovsky06a,Olmi_Del-Zanna+15a}. 
The wind magnetization parameter is equal to 1 ($\Rightarrow \sigma_0=10$),
with  a narrow ($\sim 10^\circ \Rightarrow b=10$) striped wind
region. This wind is injected into an environment corresponding to the
self-similar expanding supernova ejecta
\citep{Del-Zanna_Amato+04a}. The PWN evolution is followed for a
few hundreds of years, until it reaches a state that is self-similar,
such that results can be reasonably extrapolated to the present age of the
the Crab nebula (for example the nebular radius is set to
$\approx 5.8$ ly at a distance of $2$ kpc). The magnetic field is strongly toroidal in the inner region
close to the termination shock, becoming more tangled and developing a substantial
poloidal component in the outer region of the PWN. While the geometry of the magnetic field,  is taken
from the numerical model, its strength is left as a free parameter in
order to investigate how results change depending on the level of
equipartition between field and particles that is reached in the
nebula. Current spectral models \citep{Gelfand_Slane+09a,Bucciantini_Arons+11a,Martin_Torres+12a} suggest that the average
field in the nebula is $\sim 100$ $\mu$G, and we take this as our
fiducial value. 

Radio emitting particles are assumed to follow a power-law
distribution with energy $n(\gamma)  = n_e \gamma^{-s}$, where the
power-law index is fixed by the radio spectral index to a value
$s=1.5$  \citep{Baldwin71a,Baars72a,Bietenholz_Kassim+97a}, while the normalization $n_e$ is fixed by
requiring a spectral energy density $S_\nu = 2.5\times 10^{24}$ erg
s$^{-1}$ Hz$^{-1}$ at 10GHz \citep{Hester08a}. Given
that the radio spectrum extends without breaks down to the ionospheric
cutoff $\sim 30$ MHz, we can safely assume that the distribution
function extends as a power-law down to a minimum Lorenz factor
$\gamma_{\rm min}\ltsim 100$. Radio emitting particles are assumed to
be distributed uniformly in the nebula \citep{Olmi_Del-Zanna+14a}.

Synchrotron maps of the nebula, in the various Stoke's parameters are
computed according to standard recipes \citep{Bucciantini_del-Zanna+05a,Del-Zanna_Volpi+06a,Volpi_Del-Zanna+08b,Olmi_Del-Zanna+14a}. The local
synchrotron  emissivity is integrated along the line of sight for each point of the
plane of the sky. Given that radio emission is dominated by the
contribution of the outer parts of the PWN, where typical flow
velocity are $\ll c$, one can safely neglect Doppler boosting effects
\citep{Bucciantini17a}. The local emissivity for the Stoke's parameter $V$
describing CP is given by \citet{Legg_Westfold68a}. We include
in our modeling the presence of a small scale magnetic turbulence (named  $\sigma$, which will
not be resolved in large scale 3D models), following
the prescription by \citet{Bandiera_Petruk16a}, recently applied to X-ray modeling of PWNe
by \citet{Bucciantini_Bandiera+17a}. The assumption is that magnetic field fluctuates around
an average value $\boldsymbol{\bar{B}}$ with a Gaussian probability
distribution
\begin{align}
\mathcal{P}(\boldsymbol{B}) \propto {\rm Exp}[-(\boldsymbol{B}-\boldsymbol{\bar{B}})^2/2\sigma^2]
\end{align} 
\citet{Bandiera_Petruk16a} show that, for a power-law distribution of emitting
particles, the emissivity in $I$ and $Q$ are  simply given by the
standard synchrotron theory (corresponding to a purely ordered field, $\sigma=0$), corrected by an analytic coefficient
dependent only on the particle power-law index $s$ and the ratio
$\bar{B}^2/2\sigma^2$. Following the same approach one
can compute the correction to the emissivity in the Stoke's parameter
$V$. One finds
\begin{align}
V= V_o \left\{\Gamma\left(\frac{4+s}{4} \right)\left(
    \frac{\bar{B}_\perp}{\sqrt{2}\sigma} \right)^{-s/2}
\cdot \;_1\!F_1\left[ \frac{s}{4},1, - \frac{\bar{B}_\perp^2}{2\sigma}   \right]\right\}
\end{align}
where $V_o$ is the CP emissivity in the case $\sigma=0$ \citep{Legg_Westfold68a},
$\bar{B}_\perp$ is the average field perpendicular to the line of
sight, and corresponds to the value given by the numerical model,
and  $_1\!F_1(a,b,z)$ is the Kummer confluent hypergeometric
function. Images in the various Stoke's parameters are computed on a high resolution level where the
nebula is sampled over a $100\times 100$ pixel map, corresponding to a
resolution of $3$ arcsec. We can then convolve them with a broader
Point Spread Function (PSF) in order to model typical radio instrumental resolution,
for direct comparison with observations. The level of unresolved
turbulence $\sigma$ is chosen in order to reproduce the typical
polarized fraction observed in radio. 

\section{Results}
\label{sec:results}

In order to constrain the value of the parameter $\sigma$ describing
the level of unresolved turbulence, simulated maps of the Stoke's
parameters $I,U,Q$ computed according to the prescription by
\citet{Bandiera_Petruk16a}, were used to derive the linear polarization
properties of the Crab nebula to be directly compared with observations.  The
inclination of the nebula with respect to the plane of the sky \citep{Weisskopf_Hester+00a}
was taken into account, while we neglected the inclination on the
plane of the sky itself (the vertical direction is aligned to the axis
of the nebula, and does not correspond to the North-South direction). As a
reference we took the more recent measures by
\citet{Aumont_Conversi+10a} at 90 GHz. The simulated maps were
convolved with a PSF with a Full Width Half Maximum (FWHM) of $\sim 30\arcsec$ to match the resolution of the
observations, and regions with lower surface brightness, $I<0.05I_{\rm
  max}$, were discarded from the analysis in order to avoid spurious
edge effects from faint portions of the PWN.  We found that setting $\sigma=0.4$ the average polarized
fraction is $\sim 15\%$ with peaks up to $30\%$, in agreement with
observations.  This implies that the magnetic energy into
the turbulent unresolved component of the magnetic field $E_{\rm turb}
$, is about one half of the one into the
large scale component $E_{\rm ord}$ (the relation between the two is $E_{\rm turb} = 3\sigma^2 E_{\rm ord}$). This value is smaller than what was
found by \citet{Bucciantini_Bandiera+17a} in modeling the X-ray
surface brightness profile of the Crab torus. However, that model
assumed a fully toroidal (ordered) large scale field, while in our
3D simulation the turbulent cascade is partially resolved, at least
over the larger scales. This is compatible with the discrepancy in
the value of $\sigma$ between the two models.

\begin{figure}
	\centering
	\includegraphics[width=.5\textwidth]{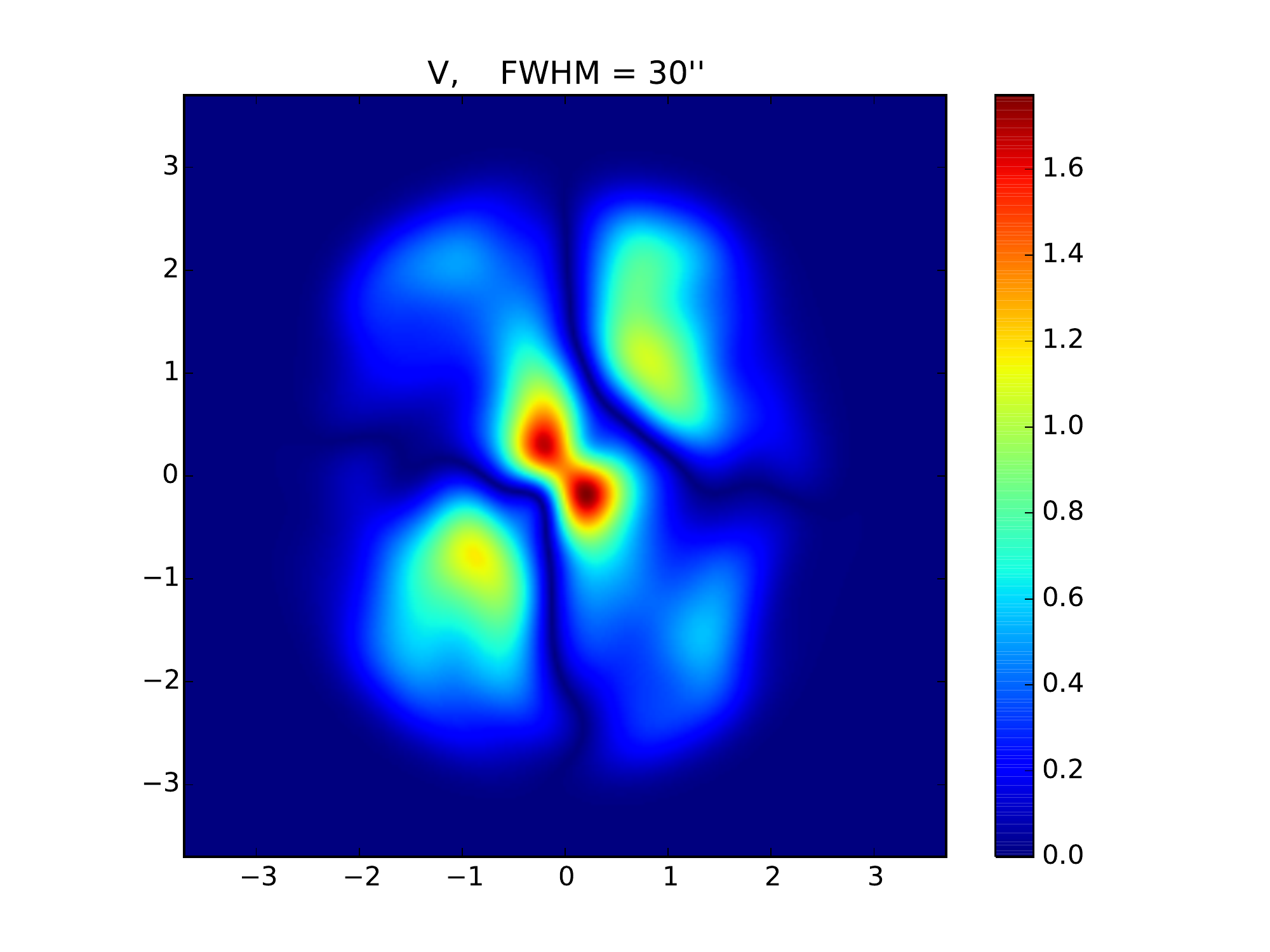}\\
	\includegraphics[width=.5\textwidth]{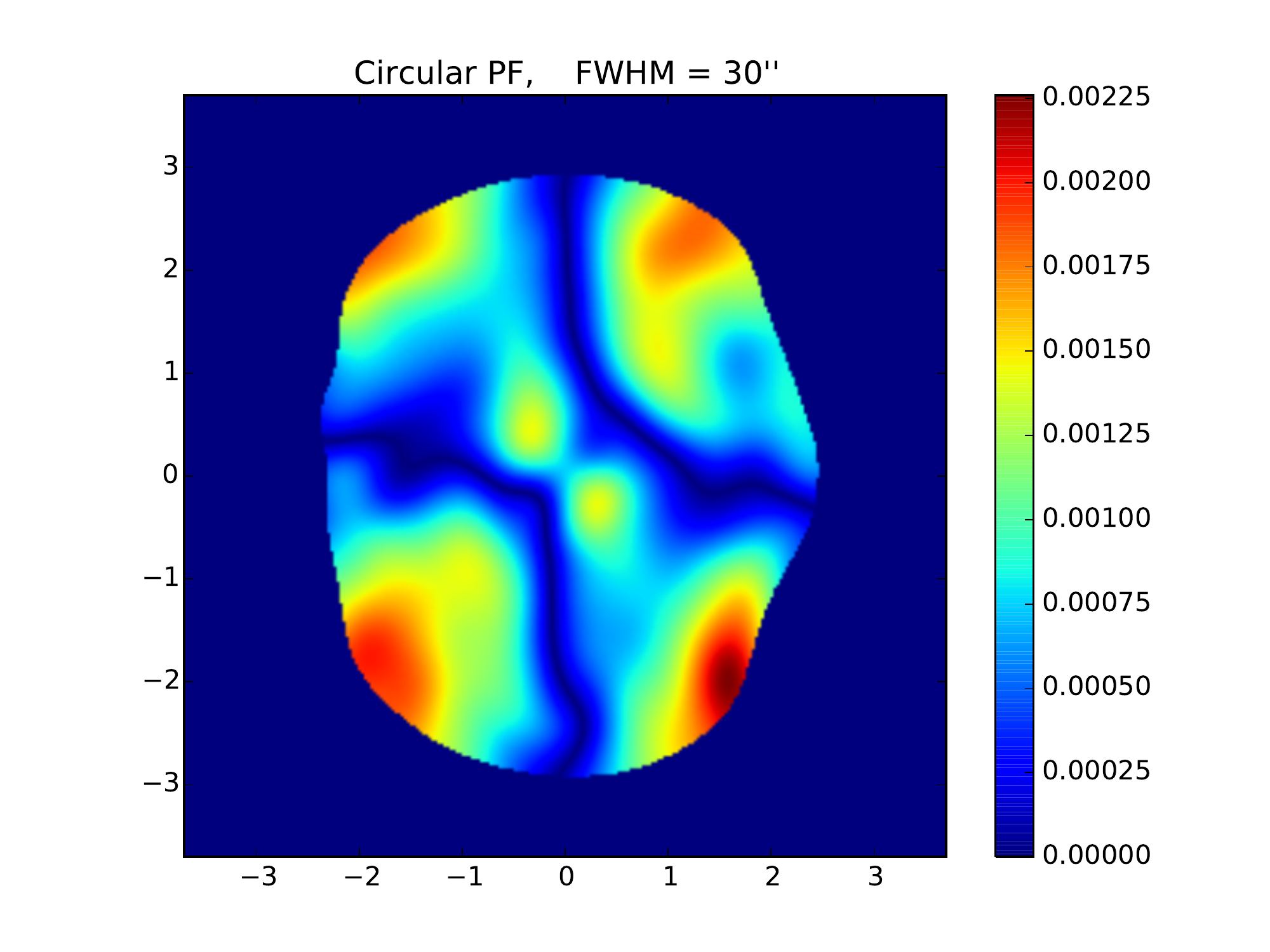}
	\caption{Upper panel: simulated map of the surface brightness of the
          Stoke's parameter $V$ at 100 MHz, for a PSF with a FWHM$=30\arcsec$, in units of
          $10^{28}$ erg s$^{-1}$ Hz$^{-1}$ str$^{-1}$. Lower panel:
          circular polarized fraction $V/I$ at 100 MHz for the same PSF. Only the
          region with $I>0.05I_{\rm max}$ is shown. Axes are in
          primes, centered on the pulsar.
     }
	\label{fig:cp1}
\end{figure}

\begin{figure}
	\centering
	\includegraphics[width=.5\textwidth]{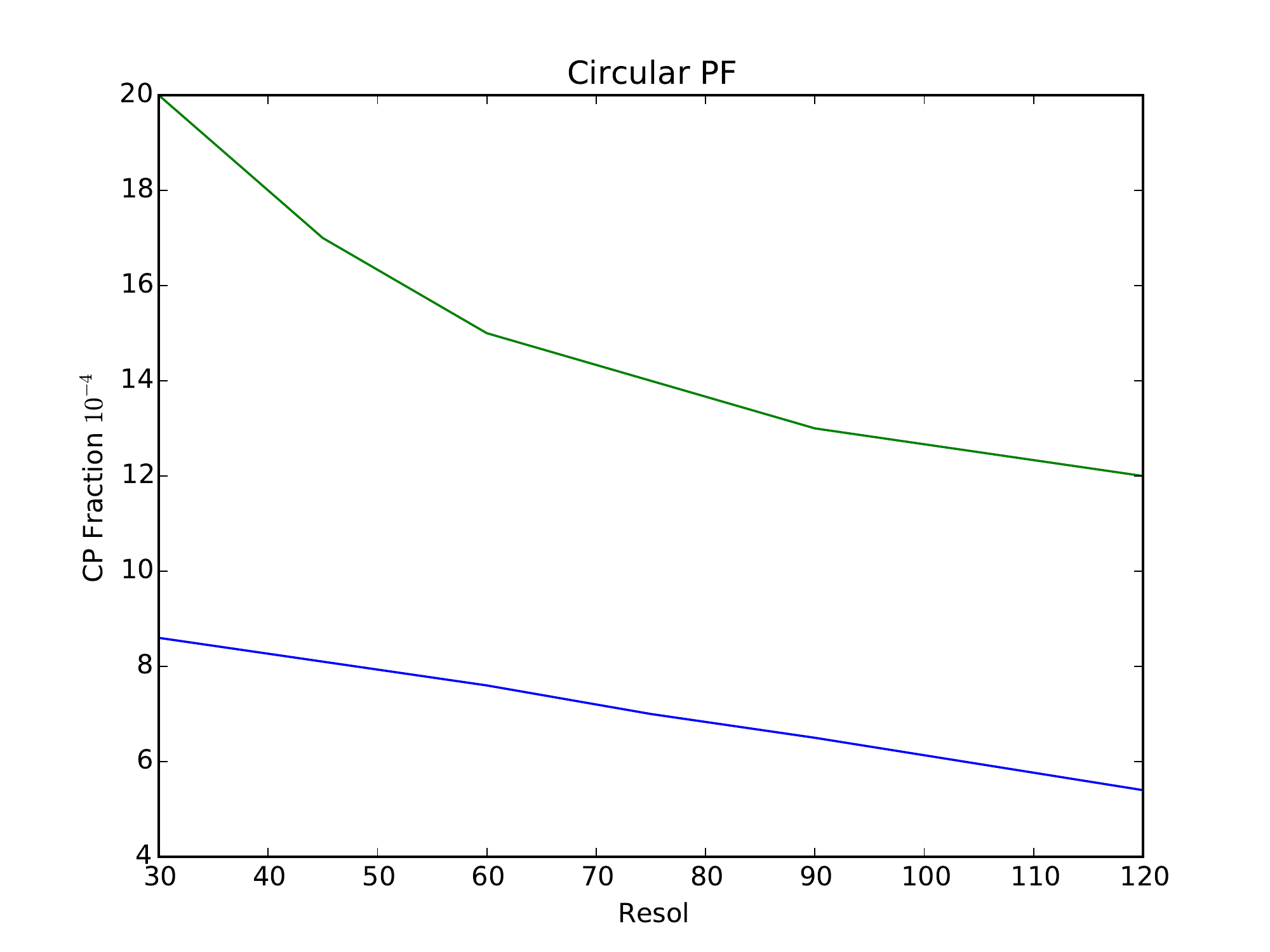}\\
	\caption{Peak circular polarized fraction (upper green curve) and
          average circular polarized fraction (lower blue curve), at
          100 MHz, from simulated maps, as a function of instrumental
          resolution given as the full
          width half maximum of the PSF in arcsec
     }
	\label{fig:cp2}
\end{figure}

In Fig.~\ref{fig:cp1} we show a map of the Stoke's parameter $V$ and
of the CP fraction computed from our 3D simulation, and convolved with
a PSF of $30\arcsec$ FWHM. Maps are shown
over the region of the nebula where the intensity is at
least 5\% of the maximum ($I<0.05I_{\rm max}$), again in order to avoid
contaminations from the faint edges. Note that $V$ and the CP fraction
scale as a function of the average field strength and observational
frequency as:  $(\langle B \rangle/100\;\mu{\rm G})^{1/2}(\nu/100{\rm \;MHz})^{-1/2}$, where the average field is defined as the square root
of the ratio of total magnetic energy (ordered large scale field plus
disordered one) over the nebular volume $\langle B \rangle=
\sqrt{8\pi (E_{\rm turb}+E_{\rm ord})/V_{\rm pwn}}$. The CP fraction
at this resolution is found to be $V/I \simeq 8.5\times10^{-4}$, with
local maxima $V/I \simeq 2\times 10^{-3}$, located however toward the
faint edges of the nebula. A more conservative estimate in the central
brighter part within $1'$ from the pulsar is $V/I\simeq 5\times 10^{-4}$. Even in the central part of the PWN, as expected from synchrotron theory, the CP
fraction is higher where the intensity tends to be smaller, and there
are large portions of the nebula where the CP fraction is below $\simeq 5\times10^{-4}$. In Fig.\ref{fig:cp2} we
show how results change with resolution both for the integrated CP
fraction and the peak CP fraction. Results clearly show  that below a
resolution  of $\sim 2^\prime$, the integrated CP fraction is halved
with respect to the value corresponding
to the result of Fig.~\ref{fig:cp1}. However the numbers are dominated
by the edges.  The high CP fraction central peaks within $1^\prime$ from the
pulsar, seen in Fig.~\ref{fig:cp1}, with CP fraction $V/I \simeq
0.0015$, disappear already at a resolution of $60\arcsec$.

Stoke's $V$ can also be generated by Faraday conversion ($I\rightarrow
V$, and $Q\rightarrow V$), either inside or
outside the source. Concerning the internal Faraday conversion, given that  radio emission extends as a power-law to the
ionospheric cutoff $\sim 30$ MHZ, one can safely assume a minimal
Lorentz factor $\gamma_{\rm min} \ltsim 100$ and use the standard analytic
formulae by \citet{Sazonov69a} for a power-law distribution \citep{Huang_Shcherbakov11a}. We
found that the induced circular polarization is a few orders of
magnitude smaller than the intrinsic one. For Faraday conversion in the
ISM we took an electron density $n_e=0.015$ cm$^{-3}$ and typical
average magnetic field $B_{\rm ISM} \approx 1\;\mu$G derived
from radio dispersion and rotation measure for the Crab pulsar and
nebula  \citep{Davidson_Terzian69a,Manchester71a,Bietenholz_Kronberg91a}. Taking a typical temperature in the range
$10^4-10^5$K, the induced CP is again found to be several orders of
magnitude smaller than the intrinsic one. 
 This implies that the computed CP maps are unaffected by propagation, and that any detection of CP
is likely to be intrinsic.

\section{Conclusions}
\label{sec:conclusion}

In this work, for the first time, we computed the expected circular
polarization from synchrotron emission in the Crab nebula, considering that the emitting plasma is mainly composed by electrons and protons, using a state of the art
3D model for the magnetic field geometry inside the PWN. Our results
clearly show that previous non detections were not constraining. The
upper limit by \citet{Wilson_Weiler97a}, obtained at a resolution of
$50"\times 150"$ FWHM, would correspond to an integrated CP fraction at
100 MHz of about $7\times10^{-4}$, which is just at the limit
predicted in Fig.~\ref{fig:cp2}, and a factor 2 above it if one
considers just the brightest part of the nebula. Even the expected
peak values at that resolution is estimated to be just a factor 2 above
their detection limit. However, as discussed previously, this is
mostly due to faint edges. The ability to identify possible peaks strongly relies
on resolution: we estimate that a PSF with a FWHM not greater than
$60\arcsec$ is required in order to see the brightest inner peaks.

The Crab nebula is a bright object, and the intensity of  the Stoke's parameter $V$
is high enough that it could be easily detected above noise, even with
short term observations, by essentially all current radio
telescopes. Unfortunately the major issue in polarization measures comes
from calibration \citep{Hales17a}. Calibration leakage among the various
Stoke's parameters and the lack of absolute calibrators set in general
limit on the CP detection with $V/I$  at best at the level of few
$10^{-3}-10^{-2}$ \citep{Tasse_van-der-Tol+13a,Perley_Butler13a,Murgia_Govoni+16a,Sokolowski_Colegate+17a}. \citet{Rayner_Norris+00a} have shown that using specifically designed
calibration strategies it is possible to get reliable measures up to $V/I\sim 2 \times
10^{-3}$ at a few GHz \citep[see also][]{Wiesemeyer_Thum+11a}, with a
typical errors $\sim$ a few $10^{-4}$, and resolution high enough to
resolve the Crab nebula inner structures. This is however again at the
limit of our predicted values, once scaled to few GHz.
Being the Crab nebula a strongly linearly
polarized source, it is not even clear  if systems with linear
polarized feeds improve on circular ones. Part of the problem stems
from the fact that there are few circularly polarized targets, and CP
science is not a strong driver in the construction of radio
facilities, even if there has been an increasing interest in CP
\citep{Perley_Butler13a,Enslin_Hutschenreuter+17a,Mao_Wang17a}. 

If special attention is devoted to construction and calibration of the next
generation radio telescopes, especially at the low frequency range, it will be feasible to reach the detection limit for the Crab nebula in the forthcoming future.

\section*{Acknowledgements}

The authors acknowledged support from the PRIN-MIUR project prot. 2015L5EE2Y "Multi-scale simulations of high-energy astrophysical plasmas".

\bibliography{Bib}{}
\bibliographystyle{mn2e}

\end{document}